\documentclass
[superscriptaddress,secnumarabic,amssymb,amsmath,nobibnotes,aps,prd,showkeys,showpacs,nofootinbib,onecolumn]{revtex4}%
\usepackage{graphicx}
\usepackage{epsf}
\usepackage{bm}
\usepackage{amsmath}
\usepackage{amsfonts}
\usepackage{amssymb}%
\providecommand{\U}[1]{\protect\rule{.1in}{.1in}}

\newcommand{\be}{\begin{equation}}
\newcommand{\ee}{\end{equation}}

\newcommand{\mincir}{\raise
-3.truept\hbox{\rlap{\hbox{$\sim$}}\raise4.truept\hbox{$<$}\ }}
\newcommand{\magcir}{\raise
-3.truept\hbox{\rlap{\hbox{$\sim$}}\raise4.truept\hbox{$>$}\ }}

\begin{document}
\title{Analytic Solution of the Starobinsky Model for Inflation}
\author{Andronikos Paliathanasis}
\email{anpaliat@phys.uoa.gr}
\affiliation{Instituto de Ciencias F\'{\i}sicas y Matem\'{a}ticas, Universidad Austral de
Chile, Valdivia, Chile}
\affiliation{Institute of Systems Science, Durban University of Technology, PO Box 1334,
Durban 4000, Republic of South Africa}

\begin{abstract}
We prove that the field equations of the Starobinsky model for inflation in a
Friedmann-Lema\^{\i}tre-Robertson-Walker constitute an integrable system as
the field equations pass the singularity test. The analytical solution in
terms of a Painlev\'{e} Series for the Starobinsky model is presented for the
case of zero and nonzero spatial curvature. In both cases the leading-order
term describes the radiation era provided by the corresponding higher-order theory.

\end{abstract}
\keywords{Cosmology; $f(R)$-gravity; Starobinsky model; Integrability}
\pacs{98.80.-k, 95.35.+d, 95.36.+x}
\maketitle
\date{\today}

\section{Introduction}

In the so-called modified/extended theories of gravity \cite{clifton} new
dynamical terms, of geometric origin, are introduced which force the evolution
of the gravitational field equations in order to explain various phenomena
which were raised by recent observations \cite{capQ,koyama}. However, in the
modified gravitational theories the new terms increase the complexity of the
field equations and even in the simplest models, such as that of an isotropic
and homogeneous universe, the existence of an analytical solution is not
obvious. Although numerical methods can be applied to approximate the
evolution of the field equations that is not sufficient for the complete study
of a theory; while the analysis of the critical points it is not sufficient to
provide us with information for the evolution of a system far from the
critical points. Consequently the existence of analytical solutions for the
field equations has lead to the application of various techniques from the
analysis of dynamical systems for the study of the integrability\footnote{For
the application of the invariant transformations in modified theories see for
instance \cite{nor1,nor2,nor3,nor4a,nor4} and references therein.}.

One of the simplest modifications of the Einstein-Hilbert Action which
consider quantum corrections is the Starobinsky model of inflation \cite{star}
with Action Integral%
\begin{equation}
S=\int d^{4}x\sqrt{-g}\left(  R+qR^{2}\right)  +\int d^{4}x\sqrt{-g}%
L_{m},\label{st.01}%
\end{equation}
where $R^{2}$ describes the quantum-gravitational effects in the early
universe and $L_{m}$ is the Lagrangian of the matter source. The latter Action
Integral corresponds to the family of the so-called quadratic theories
instance \cite{qua1,qua2,qua3}. 

The gravitational field equations are of fourth order and in the case of a
spatially flat Friedmann-Lema\^{\i}tre-Robertson-Walker (FLRW) universe with
line element\footnote{We have assumed that the lapse function in the FLRW line
element is constant, i.e., $\ N\left(  t\right)  =1$~and $a\left(  t\right)  $
denotes the scale factor.}%
\begin{equation}
ds^{2}=-dt^{2}+a^{2}\left(  t\right)  \left(  dx^{2}+dy^{2}+dz^{2}\right)
\end{equation}
are calculated to be
\begin{equation}
3\left(  \frac{\dot{a}}{a}\right)  ^{2}-54q\left(  \frac{\dot{a}}{a}\right)
^{4}+18q\left(  2\left(  \frac{\dot{a}}{a}\right)  ^{2}\frac{\ddot{a}}%
{a}-\left(  \frac{\ddot{a}}{a}\right)  ^{2}+2\left(  \frac{\dot{a}}{a}\right)
^{2}a^{\left(  3\right)  }\right)  =\rho_{m} \label{st.02}%
\end{equation}
and
\begin{equation}
2\frac{\ddot{a}}{a}+\left(  \frac{\dot{a}}{a}\right)  ^{2}+18q\left(  \left(
\frac{\dot{a}}{a}\right)  ^{4}+\frac{\ddot{a}}{a^{2}}\right)  +24q\left(
\frac{\dot{a}}{a^{2}}a^{\left(  3\right)  }-3\left(  \frac{\dot{a}}{a}\right)
^{2}\frac{\ddot{a}}{a}\right)  +12q\frac{a^{\left(  4\right)  }}{a}=-p_{m},
\label{st.04}%
\end{equation}
where $\rho_{m}$, $p_{m}$ are the energy density and the pressure of the
matter source. In the case of the vacuum the field equations (\ref{st.02}%
)-(\ref{st.04}) admit an unstable (special) nonsingular solution \cite{star}.
Moreover it is important to mention that, when the equation of state parameter
of the matter source, $\rho_{m},~p_{m}$ is that of an ideal gas, i.e.,
$p_{m}=\left(  \gamma-1\right)  \rho_{m}$, then equation (\ref{st.04}) can be
integrated to (\ref{st.02}), while in general the conservation law $\dot{\rho
}_{m}+3H\left(  \rho_{m}+p_{m}\right)  =0$ exists.

The Action Integral (\ref{st.01}) corresponds to the $f\left(  R\right)
$-theories of gravity \cite{Buda}, where $f\left(  R\right)  =R+qR^{2},$ while
a more general consideration of the Starobinsky model is the\footnote{For
reviews in $f\left(  R\right)  $-gravity see for instance
\cite{Sotiriou,odin1} while some observational constraints can be found in
\cite{oc1,oc2,oc3,oc4}.} $f\left(  R\right)  =R+qR^{n}$ theory\footnote{There
is a plethora of physical theories which has been inspired by the Starobinsky
model of inflation such as in SUGRA or in other gravitational theories, for
instance see \cite{gsta0,gsta1,gsta2,gsta3,gsta4,gsta5} and references
therein.} \cite{fr6} or \cite{fr7}. Exact solutions of power-law f(R) theories
can be found in \cite{fr8,fr9}. The importance of the Starobinsky model
(\ref{st.01}) is that it provides an inflationary scenario which is favored by
the observations \cite{planck2015}. Furthermore it has been shown that various
models of inflation are identical with the Starobinsky model when the
inflationary phase takes place \cite{keha} whereas the Mixmaster universe
provides nonchaotic trajectories \cite{riccardo}.

The introduction of a Lagrange Multiplier in $f\left(  R\right)  $-theories
\cite{capLan} can be used to reduce the order of the theory from a
fourth-order to a second-order theory by increasing at the same time the
number of degrees of freedom \cite{ccc}. In particular a new field is
introduced which is a equivalent to that of a Brans-Dicke scalar field with
zero Brans-Dicke parameter \cite{Sotiriou,defelice} the so-called O'Hanlon
theory \cite{hanlon}. Therefore the field equations in a FLRW background form
a two-dimensional canonical Hamiltonian system which describes a particle
moving in a flat space while the potential which forces the evolution of the
system is related with the form of the $f\left(  R\right)  $ function. Because
the scalar field description is that of Brans-Dicke field the theory is
defined in the Jordan frame. Hence under a conformal transformation a
minimally coupled scalar field is defined and the theory is defined now in the
Einstein frame\footnote{For a discussion between these two frames see
\cite{fram1,fram2} and references therein.}. Therefore the Starobinsky model
can be seen as a mechanics to provide a minimally scalar field \cite{barcots}
to drives the inflationary phase of the universe, for a review see
\cite{bamba1}

By using the property that the field equations describe a canonical
Hamiltonian system various functions $f\left(  R\right)  $ have been
determined in which the field equations admit conservation laws which are
linear or quadratic in the momentum \cite{nor4}, while recently in \cite{apla}
it was found that the cosmological model $f\left(  R\right)  =R+qR^{n}$
\ passes the singularity test and is integrable for some values of the power
$n$. However, the case $n=2$, which is that of the Starobinsky model has been
ruled out and the main reason is that for $n\neq2$ the field equations admit
singular special solutions following from the the $R^{n}$ term; in contrast to
the Starobinsky model in which the $R^{2}$ term provides a nonsingular
solution as mentioned above.

A specific $f\left(  R\right)  $ theory provides a de Sitter universe if there
exists $R=R_{0}$ \ such that the Barrow-Ottowill \cite{BarOtt} condition holds%
\begin{equation}
R_{0}f^{\prime}\left(  R_{0}\right)  -2f\left(  R_{0}\right)  =0.
\label{fr.26a}%
\end{equation}
\qquad It is straightforward to see that for arbitrary $R_{0}$, that is,
$R_{0}\rightarrow R$, the latter condition can be seen as a first-order
differential equation with solution the quadratic function $f\left(  R\right)
=f_{0}R^{2}$, where $f_{0}$ is a constant of integration.

\section{Integrability of the field equations}

In the case of the vacuum the field equations in $f\left(  R\right)  =R^{2}$
theory are%
\begin{equation}
2a^{2}\dot{a}a^{\left(  3\right)  }-3\left(  \dot{a}\right)  ^{4}-a\ddot
{a}\left(  a\ddot{a}-2\dot{a}^{2}\right)  =0\label{st.05}%
\end{equation}
and
\begin{equation}
2a^{3}a^{\left(  4\right)  }+4a^{2}\dot{a}a^{\left(  3\right)  }+3\left(
\dot{a}\right)  ^{4}+3a\ddot{a}\left(  a\ddot{a}-4\left(  \dot{a}\right)
^{2}\right)  =0.\label{st.06}%
\end{equation}
The two equations are not independent and derivation of (\ref{st.05}) gives
the fourth-order equation (\ref{st.06}). There are various ways in which
equation (\ref{st.05}) can be written as a first-order ordinary differential
equation\footnote{The field equations (\ref{st.05}), (\ref{st.06}) admit as
point symmetries the $\partial_{t},~t\partial_{t}~$and $a\partial_{a}$ vector
fields which form the $\left\{  2A_{1}\otimes A_{1}\right\}  $ Lie algebra.}.
If we select the new dependent variable $w=\frac{1}{u}\frac{du}{dv}$ and
independent variable $v$, where $u=\dot{a},~v=a$ then equation (\ref{st.05})
becomes the following Riccati equation%
\begin{equation}
2\frac{dw}{dv}+3w^{2}+2\frac{w}{v}-\frac{3}{v^{2}}=0\label{st.07}%
\end{equation}
with solution $w\left(  v\right)  =\frac{v^{3}-w_{0}}{v\left(  v^{3}%
+w_{0}\right)  }$, where $w_{0}$ is a constant of integration. Therefore it
follows that $\frac{H\left(  t\right)  }{H_{0}}=\left(  v_{0}^{2}a^{-\frac
{3}{2}}+a^{\frac{3}{2}}\right)  ^{\frac{2}{3}}$, where $H\left(  t\right)
=\frac{\dot{a}}{a}$, and for initial conditions such that $v_{0}=0$ provides
the closed-form solution~$a\left(  t\right)  \simeq t^{-1}$. That is not the
unique case. In order to see that consider now the new variables $\left\{
x,y\right\}  =\left\{  H\left(  t\right)  ,\frac{d}{dt}\left(  H\left(
t\right)  \right)  \right\}  $ equation (\ref{st.05}) equation (\ref{st.05})
takes the form of the linear equation~%
\begin{equation}
2\frac{dy}{dx}-y+6x^{2}=0\label{st.08}%
\end{equation}
~\ with solution~$y\left(  x\right)  =-2x^{2}+u_{1}\sqrt{x}~$, that is
$\int\frac{dH}{u_{1}\sqrt{H}-2a^{2}}=\left(  t-t_{0}\right)  $ where in the
limit $u_{1}=0$, gives $H\left(  t\right)  =\frac{1}{2\left(  t-t_{0}\right)
}$, that is, $a\left(  t\right)  =a_{0}\sqrt{\left(  t-t_{0}\right)  }$, which
is an ideal gas solution which mimics radiation solution, while it is a
singular (special) solution. This singular solution is used below in order to
prove the integrability of the Starobinsky model. The existence of the
radiation solution it is not a surprise in the sense that $f\left(  R\right)
$-gravity can provides always a radiation epoch in the evolution of the
universe \cite{amrad}. However the radiation solution have been investigated
before in a higher-order theory which include the Starobinsky term as also
other terms follows from the Gauss-Bonnet invariant in \cite{cots1,cots2,midd}%
. Moreover the radiation solution in quadratic theories has been found that
can describes a past isotropic singularity for the Bianchi I universe
\cite{hervik}.

The method that we apply is that of the singularity analysis and specifically
we follow the ARS algorithm \cite{Abl1,Abl2,Abl3}.\ Singularity analysis is a
powerful method which has been applied in cosmological studies for the
reconstruction of the analytical solution of various models
\cite{apla,Cotsakis,pb,ftAn,ksAn}. We omit the properties of the singularity
analysis and we refer the reader to the extended review \cite{buntis}.

We continue by firstly applying the method for the quadratic theory $f\left(
R\right)  =R^{2}$ and consider now equation (\ref{st.06}).

We find that the leading-order behavior is the power-law solution $a\left(
t\right)  =a_{0}\tau^{1/2}$,~where $\tau=t-t_{0}$ and $t_{0}$ denotes the
position of the singularity. The application of the ARS algorithm provides the
resonances to be $s_{1}=-1~,~s_{2}=0~,~s_{3}=\frac{3}{2}~$and $s_{4}=\frac
{5}{2},$which means that the analytic solution is expressed by the Right
Painlev\'{e} Series \cite{Feix97}%
\begin{equation}
a\left(  t\right)  =a_{0}\tau^{\frac{1}{2}}+a_{1}\tau+a_{2}\tau^{\frac{3}{2}%
}+a_{3}\tau^{2}+%
{\displaystyle\sum\limits_{i=4}^{\infty}}
a_{4}\tau^{\frac{1+i}{2}}, \label{st.09}%
\end{equation}
where the constants of integration are the $a_{0},~a_{3},~a_{5}$ and the
position of the singularity $t_{0}$. However, with the use of (\ref{st.05}) we
find that $a_{5}=0$, while the calculation of the first coefficient constants
gives the solution to be
\begin{equation}
a\left(  t\right)  =a_{0}\tau^{\frac{1}{2}}+a_{3}\tau^{2}+\frac{19}{32}%
\frac{\left(  a_{3}\right)  ^{2}}{a_{0}}\tau^{\frac{7}{2}}+\frac{17}{264}%
\frac{\left(  a_{3}\right)  ^{3}}{\left(  a_{0}\right)  ^{2}}\tau^{5}+%
{\displaystyle\sum\limits_{j=10}^{\infty}}
a_{j}\tau^{\frac{1+j}{2}}. \label{st.10}%
\end{equation}

For the field equations of the Starobinsky model we apply the same algorithm
and we find the same resonances as those of the quadratic model, which means
that the analytic solution is given by expression (\ref{st.09}) or
specifically by calculation the first nine coefficient constants the solution
is%
\begin{align}
a\left(  t\right)   &  =a_{0}\tau^{\frac{1}{2}}+a_{3}\tau^{2}-\frac{a_{0}%
}{72q}\tau^{\frac{5}{2}}+a_{5}\tau^{3}+\frac{19}{32}\frac{\left(
a_{3}\right)  ^{2}}{a_{0}}\tau^{\frac{7}{2}}-\frac{5}{252}\frac{a_{3}}{q}%
\tau^{4}+\nonumber\\
&  +\left(  \frac{41}{259200q^{2}}+\frac{a_{3}a_{5}}{16a_{0}}\right)
\tau^{\frac{9}{2}}+\left(  \frac{17}{264}\frac{\left(  a_{3}\right)  ^{3}%
}{\left(  a_{0}\right)  ^{2}}-\frac{a_{5}}{132q}\right)  \tau^{5}+%
{\displaystyle\sum\limits_{j=10}^{\infty}}
\bar{a}_{j}\tau^{\frac{1+j}{2}}, \label{st.11}%
\end{align}
where the constants of integration are again the coefficients $a_{0}%
,~a_{3},~a_{5}$ and the position of the singularity $t_{0}$, while the
constraint equation (\ref{st.02}) gives that $a_{5}=0$ or, if we assume the
existence of a dust fluid, that is,~$p_{m}=0$ and $\rho_{m}=\rho_{m0}a^{-3}$,
it follows that $\rho_{m0}=\frac{315}{2}qa_{5}\left(  a_{0}\right)  ^{2}$.
While in the latter scenario it is important to mention that the term
$a_{0}t^{1/2}~$describes the leading-order behaviour.

From the values of the resonances it is easy to see that the radiation
solution is an unstable solution\footnote{For a discussion on the relation
between the values of the resonances and the stability of the leading-order
behaviour see \cite{leachss}.}, while the field equations of the Starobinsky
model for inflation in a spatially flat FLRW spacetime pass the singularity
test and are integrable.

\section{Discussion}

Singularity analysis is a powerful method to study the integrability of
dynamical systems. However, it has a basic disadvantage in that it is
coordinate dependent. That is the reason that the Starobinsky model did not
pass the singularity analysis in the consideration of \cite{apla}. The reason
is that in the space of variables $\left\{  a,R\right\}  $, in which usually
the $f\left(  R\right)  $-gravity is referred, the leading-order behaviour
$a\left(  t\right)  =a_{0}t^{\frac{1}{2}}$, provides a singular behaviour for
only one of the dynamical variables while for the Ricci Scalar it is a
constant. However, we overpassed that problem by working directly on the
fourth-order differential equation and without using the Lagrange Multiplier.

We now consider the case of nonzero spatially curved spacetime. Hence for the
Action Integral (\ref{st.01}) the field equations are derived to be
\begin{align}
\frac{\rho_{m0}}{a^{3}}  &  =3\left(  \frac{\dot{a}}{a}\right)  ^{2}%
-54q\left(  \frac{\dot{a}}{a}\right)  ^{4}+18q\left(  2\left(  \frac{\dot{a}%
}{a}\right)  ^{2}\frac{\ddot{a}}{a}-\left(  \frac{\ddot{a}}{a}\right)
^{2}+2\left(  \frac{\dot{a}}{a}\right)  ^{2}a^{\left(  3\right)  }\right)
+\nonumber\\
&  ~~~~~~~~~~+\frac{k}{2a^{2}}-6\frac{qk}{a^{2}}\left(  \frac{\ddot{a}}%
{a}\right)  +\frac{qk^{2}}{a^{4}} \label{st.12}%
\end{align}%
\begin{align}
0  &  =2\frac{\ddot{a}}{a}+\left(  \frac{\dot{a}}{a}\right)  ^{2}+18q\left(
\left(  \frac{\dot{a}}{a}\right)  ^{4}+\frac{\ddot{a}}{a^{2}}\right)
+24q\left(  \frac{\dot{a}}{a^{2}}a^{\left(  3\right)  }-3\left(  \frac{\dot
{a}}{a}\right)  ^{2}\frac{\ddot{a}}{a}\right)  +\nonumber\\
&  ~~~~~~~~~~~~~~+12q\frac{a^{\left(  4\right)  }}{a}+\frac{k}{6a^{2}}%
+2\frac{qk}{a^{2}}\left(  \left(  \frac{\dot{a}}{a}\right)  ^{2}-2\frac
{\ddot{a}}{a}\right)  -\frac{qk^{2}}{6a^{4}}, \label{st.13}%
\end{align}
where for the matter source we assumed that of a dust fluid.

We apply the ARS algorithm and we find that the solution is expressed again by
the Right Painlev\'{e} Series\ (\ref{st.10}) where now the coefficient
constants depend also upon the curvature $k$. For instance the first terms of
the solution are
\begin{equation}
a\left(  t\right)  =a_{0}\tau^{\frac{1}{2}}-\frac{k}{12a_{0}}\tau^{\frac{3}%
{2}}+a_{3}\tau^{2}-\left(  \frac{a_{0}}{72q}+\frac{k^{2}}{288\left(
a_{0}\right)  ^{3}}\right)  \tau^{\frac{5}{2}}+a_{5}\tau^{3}+%
{\displaystyle\sum\limits_{r=6}^{\infty}}
\bar{a}_{j}\tau^{\frac{1+r}{2}} \label{st.14}%
\end{equation}
where from (\ref{st.12}) follows $\rho_{m0}=\frac{315}{2}qa_{5}\left(
a_{0}\right)  ^{2}+30a_{3}qk.$

We conclude that the Starobinsky model for inflation in a FLRW spacetime with
or without spatial curvature it is an integrable system. Last but not least
from the singularity analysis we found that the radiation era is described by
a unstable point which is in agreement with the dynamical analysis for a
higher-order theory \cite{cots1,cots2}.

\begin{acknowledgments}
The author acknowledges financial support of FONDECYT grant no. 3160121 and
thanks the Durban University of Technology for the hospitality provided while
part of this work was performed.
\end{acknowledgments}

\end{document}